\documentclass[aps,twocolumn,superscriptaddress]{revtex4}

\usepackage{graphicx}
\usepackage{amsmath}
\usepackage{amssymb}

\newcommand{\unites}[1]{\; \mathrm{#1}}   

\begin{document}
\title{Capturing photoelectron motion with guiding fictitious particles}

\author{J. Dubois}
\affiliation{Aix Marseille Univ, CNRS, Centrale Marseille, I2M, Marseille, France}
\author{S. A. Berman}
\affiliation{Aix Marseille Univ, CNRS, Centrale Marseille, I2M, Marseille, France}
\affiliation{School of Physics, Georgia Institute of Technology, Atlanta, Georgia 30332-0430, USA}
\author{C. Chandre}
\affiliation{Aix Marseille Univ, CNRS, Centrale Marseille, I2M, Marseille, France}
\author{T. Uzer}
\affiliation{School of Physics, Georgia Institute of Technology, Atlanta, Georgia 30332-0430, USA}

\begin{abstract}
Photoelectron momentum distributions (PMDs) from atoms and molecules undergo qualitative changes as laser parameters are varied. We present a model to interpret the shape of the PMDs. The electron's motion is guided by a fictitious particle in our model, clearly characterizing two distinct dynamical behaviors: direct ionization and rescattering. As laser ellipticity is varied, our model reproduces the bifurcation in the PMDs seen in experiments. 
\end{abstract}


\maketitle

Subjecting atoms or molecules to intense laser fields gives rise to a variety of non-perturbative and highly nonlinear phenomena, such as high-harmonic generation (HHG), non-sequential multiple ionization (NSMI), and high-order above-threshold ionization (ATI). All these phenomena are based on the key mechanism of attosecond physics, namely the recollision~\citep{Corkum1993, Schafer1993, Corkum2007,  Agostini2008, Krausz2009, Becker2012}. A recollision is obtained when (i) an electron tunnel-ionizes, (ii) freely travels in the laser field, and then upon return to the ionic core, (iii) either recombines into an atomic or molecular bound state, or undergoes inelastic or elastic scattering. Ionized electron rescattering has broad applications in atomic and molecular physics. By experiencing a strong ion-electron interaction, rescattered electrons probe the atomic or molecular structure. This is the basis for imaging techniques, e.g., laser-induced electron diffraction~\citep{Zuo1996, Meckel2008, Peters2011} (LIED) for molecular imaging~\citep{Blaga2012} and photoelectron holography~\citep{Huismans2011}. These techniques exploit the fact that photoelectron momentum distributions (PMDs) encode information on the structure of the atom or the molecule. Understanding the photoelectron dynamics and identifying the mechanisms responsible for the shape of the PMDs is an essential step towards predicting and controlling~\citep{Kerbstadt2017} these strong-field phenomena.

\begin{figure}
\centering
\includegraphics[width=0.5\textwidth]{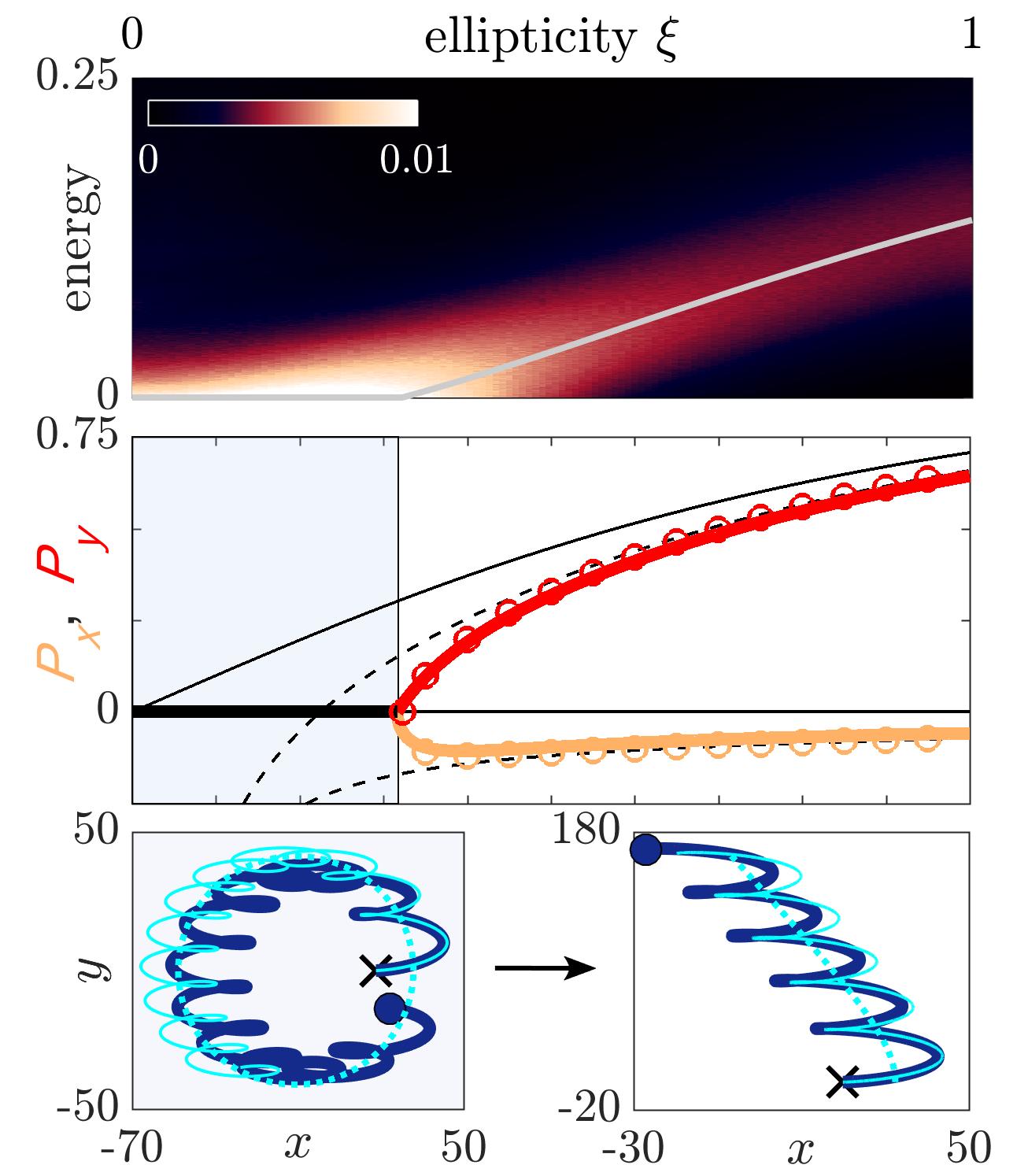}
\caption{Upper panel: ATI spectrum as a function of ellipticity, computed using CTMC from Hamiltonian~\eqref{eq:main_Hamiltonian}. The color scale indicates the probability distribution of photoelectron energies. The grey curve is the prediction of our model. Middle panel: The T-trajectory (see text) final momentum $\mathbf{P} = P_x \hat{\mathbf{x}} + P_y \hat{\mathbf{y}}$ as a function of the laser ellipticity, scaled by $E_0/\omega$. The solid colored curves and circles are computed using our model and Hamiltonian~\eqref{eq:main_Hamiltonian}, respectively. The shaded area is where the T-trajectory is rescattered in both our model and Hamiltonian~\eqref{eq:main_Hamiltonian}. The solid and dashed black curves are computed using the SFA and the perturbed SFA, respectively (the lower curves correspond to $P_x$ and the upper curves to $P_y$). Lower panels: The T-trajectory for $\xi = 0.25$ (left panel) and $\xi = 0.5$ (right panel) shown for $f=1$. The blue and cyan solid curves are the T-trajectory of Hamiltonian~\eqref{eq:main_Hamiltonian} and our model, respectively. The dashed cyan curve is the fictitious particle trajectory guiding the T-trajectory. The black crosses show the initial position of the T-trajectory. The axes use atomic units unless stated otherwise.}
\label{fig:figure1}
\end{figure}

\par
As laser parameters are varied, the shape of the PMDs undergoes drastic changes. To assess these qualitative changes in experiments~\citep{Landsman2013, Li2017}, the location of the peaks of the PMDs are followed as a function of the laser ellipticity. For low ellipticity, the shape of the PMD is a single cloud peaked at the origin, as a signature of Coulomb focusing~\citep{Brabec1996, Comtois2005}. For larger ellipticities, the cloud splits into two lobes as the Coulomb focusing recedes. Along the major polarization axis, the lobes' peaks are shifted from the origin, which is a signature of Coulomb asymmetry~\citep{Goreslavski2004, Bandrauk2000}. The hypothesis made in Ref.~\citep{Landsman2013} is that there is a bifurcation when varying the ellipticity of the laser field. This bifurcation translates into a bifurcation in the ATI spectrum --that is, the energy distribution of the ionized electrons--, as observed in the upper panel of Fig.~\ref{fig:figure1}. When the peak of the PMD is near the origin, the maximum of the ATI spectrum is near zero energy. When the PMD splits into two lobes, the energy at which the ATI is maximum increases (mostly linearly) with increasing ellipticity.
\par
Both classical~\citep{Li2013} and quantum~\citep{Li2014} simulations successfully reproduce the PMDs observed experimentally. However, the underlying dynamical mechanism leading to the drastic changes of shape of these distributions for varying ellipticities is an open question. Standard and widely used methods for the interpretation of the PMDs, like the strong-field approximation~\citep{Corkum1993,Schafer1993} (SFA) and the Coulomb-perturbed SFA~\citep{Goreslavski2004}, fail to predict these changes at low ellipticities, in particular the bifurcation observed in Ref.~\citep{Landsman2013}. The SFA neglects the Coulomb field after tunnel-ionization so it cannot capture the Coulomb asymmetry, and the perturbative treatment of the SFA is not sufficient to capture well Coulomb focusing. Our objective in this Letter is to explain the PMDs and their qualitative changes in terms of microscopic mechanisms given by the electron dynamics as laser parameters are varied, using a method which fully takes into account the Coulomb field.
\par
We begin by building a reduced classical model which reproduces the PMDs and which clearly exhibits the bifurcation in question. Analyzing this model in terms of its trajectories allows us to uncover the mechanisms responsible for the bifurcation. In a nutshell, we demonstrate that the bifurcation of the ATI spectrum is a consequence of the depopulation of the Rydberg states of the guiding fictitious particle after a critical ellipticity. The simplicity of our reduced model allows us to obtain an explicit expression for the critical ellipticity as a function of the parameters of the laser and the atom. With our model in hand, we can predict the shape of the PMDs, thereby providing essential information for imaging techniques.


\par
Here, our framework is the three-step model. In step one of the recollision process, the electron tunnel-ionizes through the barrier induced by the laser field on the ionic core potential~\citep{Keldysh1965,Ammosov1986}. We consider an elliptically polarized electric field $\mathbf{E}(t) = f(t) E_0 /\sqrt{\xi^2+1} [  \cos (\omega t) \hat{\mathbf{x}} + \xi  \sin (\omega t) \hat{\mathbf{y}}]$, where $E_0$, $\omega$, $f$ and $\xi$ are the field amplitude, frequency, envelope and polarization, respectively. After tunnel-ionization, the initial conditions of the electron $(\mathbf{r}_0, \mathbf{p}_0, t_0)$ are determined by $t_0$ and $p_{\perp}$, the ionization time and the initial transverse velocity, respectively. The electron is initially at the outer edge of the potential barrier, in the opposite direction of the electric field, i.e., $\mathbf{r}_0 = - [ I_p \mathbf{E}(t_0)/2|\mathbf{E}(t_0)|^2] [ 1+ ( 1 - 4|\mathbf{E}(t_0)|/I_p^2 )^{1/2}]$, where $I_p$ is the ionization potential of the atom. The initial longitudinal velocity of the electron is zero, i.e., $\mathbf{p}_0 = p_{\perp} \hat{\mathbf{n}}$, for a unit vector $\hat{\mathbf{n}}$ such that $\hat{\mathbf{n}} \cdot \mathbf{E}(t_0) = 0 $. In step two, the trajectory of the electron is obtained classically. In classical trajectory Monte Carlo (CTMC) simulations, ensembles of trajectories are integrated, with each one weighted by the Ammosov-Delone-Krainov~\citep{Ammosov1986} (ADK) ionization rate corresponding to the trajectory's $t_0$ and $p_{\perp}$. The trajectory with the highest weight corresponds to the trajectory initiated with zero velocity ($p_{\perp}=0$) at the peak of the electric field, when the barrier width is the thinnest. We refer to this trajectory as the T-trajectory. Here, we take the ionization time of the T-trajectory to be $\omega t_0 = \pi$. The final momentum of the T-trajectory is denoted $\mathbf{P} = P_x \hat{\mathbf{x}} + P_y \hat{\mathbf{y}}$. We assume that when the T-trajectory is not rescattered, the location of the peak of the PMDs is at $\mathbf{P}$. 

\par
In the SFA, the T-trajectory reaches the detector without experiencing a recollision with the ionic core for all laser polarizations, with a final momentum equal to its initial drift momentum. In the middle panel of Fig.~\ref{fig:figure1}, we show the final momentum of the T-trajectory, which in the SFA is $\mathbf{P}^{\mathrm{SFA}} =  \hat{\mathbf{y}}  (E_0/\omega) \xi/\sqrt{\xi^2+1}$. The SFA solution does not exhibit a bifurcation for increasing ellipticity, in contradiction with the ATI spectrum depicted in the upper panel of Fig.~\ref{fig:figure1} and the experimental results~\citep{Landsman2013, Li2017}.

\par
In order to remedy this shortcoming, a Coulomb-perturbed SFA~\citep{Goreslavski2004} is used in Ref.~\citep{Landsman2013}. The correction of the final electron momentum is given by $\Delta \mathbf{P} = - \int_{t_0}^{\infty} \mathbf{r}^{\mathrm{SFA}}(t)/|\mathbf{r}^{\mathrm{SFA}}(t)|^3 \mathrm{d}t$, where $\mathbf{r}^{\mathrm{SFA}} (t)$ is the SFA electron trajectory. In the middle panel of Fig.~\ref{fig:figure1}, we see that the Coulomb-corrected final momentum of the T-trajectory, i.e., $\mathbf{P} \approx \mathbf{P}^{\mathrm{SFA}} + \Delta \mathbf{P}$, does not exhibit a bifurcation for increasing ellipticity either, nor does it predict a change of dynamical behavior of the T-trajectory. In addition, it was noted in Ref.~\citep{Landsman2013} that this method does not predict correctly the location of the center of the PMDs for low ellipticities both in $P_x$ and in $P_y$. Hence, for low ellipticities and this range of laser parameters, a perturbed SFA is not the adapted framework for including the Coulomb interaction in order to assess the PMDs.

\par
Instead of perturbing the SFA, we consider here an averaging method over a fast timescale~\citep{Cary1983} to describe the photoelectron dynamics. In the dipole approximation formulated in length gauge, the dynamics of the electron interacting with an electric field and an ionic core  is governed by Hamiltonian
\begin{equation}
\label{eq:main_Hamiltonian}
H (\mathbf{r}, \mathbf{p}, t) = \dfrac{|\mathbf{p}|^2}{2} + V(\mathbf{r}) + \mathbf{r} \cdot \mathbf{E}(t) ,
\end{equation}
where atomic units (a.u.) are used unless stated otherwise. Here, the atom is $\mathrm{He}$, the field wavelength is $\lambda = 780\unites{nm}$ ($\omega = 0.0584 \unites{a.u.}$) and the laser intensity is $I = 8\times 10^{13} \unites{W \cdot cm}^{-2}$ ($E_0 = 0.048 \unites{a.u.}$). The field envelope $f$ consists of a two laser-cycle plateau followed by a two laser-cycle linear ramp-down, unless stated otherwise. The position of the electron is $\mathbf{r}$, and its canonically conjugate momentum is $\mathbf{p}$. We use a soft Coulomb potential~\citep{Javanainen1988} $V(\mathbf{r}) = - (|\mathbf{r}|^2 + 1)^{-1/2}$ to describe the ion-electron interaction.


\par
Averaging Hamiltonian~\eqref{eq:main_Hamiltonian} over the fast timescale, set by the period of the laser field, using Lie transform perturbation theory~\citep{Cary1983} reveals that a fictitious particle guides the electron, as shown in the lower panels of Fig.~\ref{fig:figure1}. We build a hierarchy of models using an iterative method for computing the $n$-th order perturbative expansion from the $(n-1)$-th one with Lie transforms. At the lowest order of the perturbative expansion, the electron phase space coordinates are of the form
\begin{subequations}
\label{eq:change_coordinates}
\begin{eqnarray}
\mathbf{r} &=& \mathbf{r}_g + \mathbf{E}(t)/\omega^2 , \\
\mathbf{p} &=& \mathbf{p}_g + \mathbf{A} (t) ,
\end{eqnarray}
\end{subequations}
where $(\mathbf{r}_g, \mathbf{p}_g)$ are the canonically conjugate variables of the guiding fictitious particle, and $\mathbf{A}(t)$ is the vector potential. Here, it is straightforward to see that $\mathbf{p}_g$ is the electron drift-momentum. The guiding fictitious particle dynamics is governed by the averaged Hamiltonian
\begin{equation}
\label{eq:averaged_Hamiltonian_H2}
\overline{H} (\mathbf{r}_g, \mathbf{p}_g) = \dfrac{|\mathbf{p}_g|^2}{2} + V_{\mathrm{eff}}(\mathbf{r}_g, \mathbf{p}_g) .
\end{equation}
We notice that this Hamiltonian no longer depends on time, as a result of averaging. Consequently, its energy $\mathcal{E} = \overline{H}(\mathbf{r}_g, \mathbf{p}_g)$ is conserved. At the lowest order in the perturbative expansion, $V_{\mathrm{eff}} (\mathbf{r}_g, \mathbf{p}_g) = V (\mathbf{r}_g)$. Thus, the angular momentum of the guiding fictitious particle is also conserved, and the system is integrable in the Liouville sense. At higher order in the perturbative expansion, the effective potential corresponds to the first nontrivial order of the Kramers-Henneberger potential~\citep{Bhatt1988}, and depends on the laser parameters. In this case, the angular momentum is no longer conserved, and as a consequence, the averaged system is no longer integrable. Our reduced model is valid for all positive energies and when $\omega \gg \omega_g = (2|\mathcal{E}|)^{3/2}$ for negative energy, where $\omega_g$ is the approximate frequency of the guiding fictitious particle trajectory, provided the electron and the guiding fictitious particle are outside the ionic core region. Here, we focus on the lowest-order model. 


\begin{figure}
\centering
\includegraphics[width=0.5\textwidth]{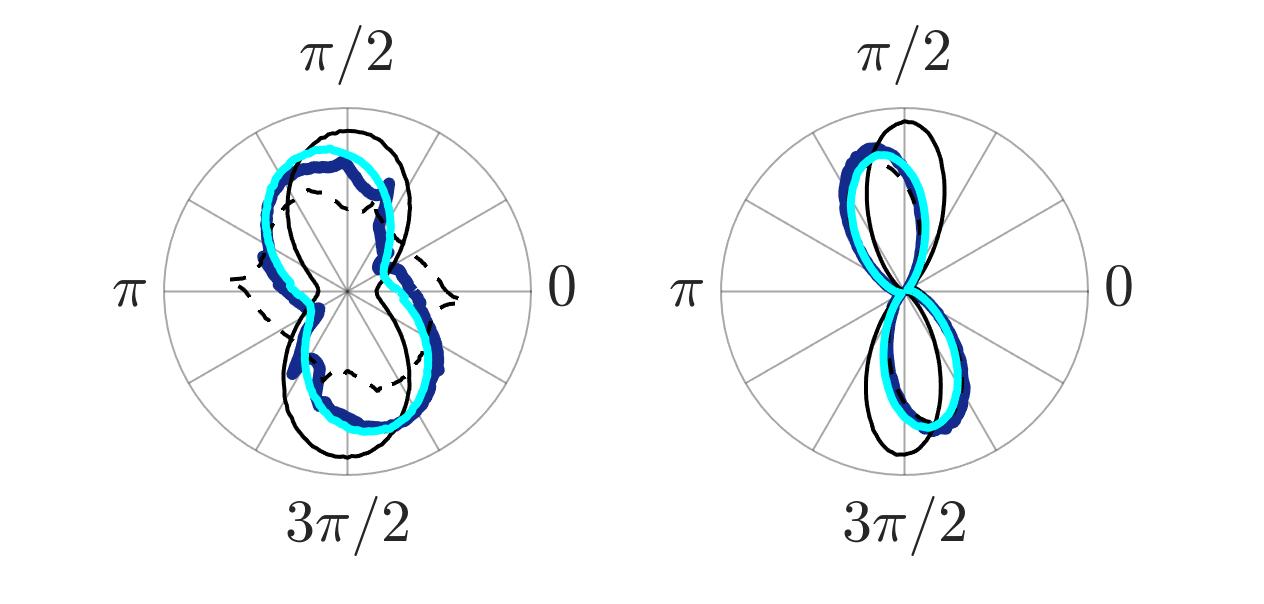}
\caption{Polar plots of the PADs for $\xi = 0.25$ (left panel) and $\xi = 0.5$ (right panel) computed using CTMC. The computation from our model (cyan) is in agreement with the one from Hamiltonian~\eqref{eq:main_Hamiltonian} (blue). Also shown are computations using the SFA (solid black line) and the perturbed SFA (dashed black line).}
\label{fig:figure2}
\end{figure}

\par
In our model, after tunnel-ionization, the electron is driven by a guiding fictitious particle. The initial conditions of the fictitious particle guiding the electron are determined by substituting the initial conditions of the electron $(\mathbf{r}_0, \mathbf{p}_0, t_0)$ in Eqs.~\eqref{eq:change_coordinates}. Then, the guiding fictitious particle dynamics is governed by Hamiltonian~\eqref{eq:averaged_Hamiltonian_H2}. When the electric field is turned off, the vector potential vanishes and the electron coordinates, in particular the momenta, become the same as that of the guiding fictitious particle. Figure~\ref{fig:figure2} shows photoelectron angular distributions (PADs) computed using CTMC methods from Hamiltonian~\eqref{eq:main_Hamiltonian}, which is compared with CTMC from the SFA, the perturbed SFA~\citep{Goreslavski2004} and our model. In the left panel, we observe excellent agreement between the prediction of our model and the full system before averaging. Moreover, since only direct electrons, i.e., the ones that do not undergo rescattering, reach the detector in our model, it becomes possible to locate rescattered electron contributions in the full system. For example, we observe two peaks around $\pi/3$ and $4\pi/3$ in the CTMC curve that are absent in the PAD of our model, corresponding to the rescattered electron contribution. In the right panel, we observe that predictions of both the perturbed SFA and our model are in agreement with the full system. The shift of the true PADs compared to the SFA prediction is known as the Coulomb asymmetry~\citep{Bandrauk2000, Goreslavski2004}. In order to understand this phenomenon from a dynamical point of view, as well as the bifurcation in the ATI spectrum, we apply our model to analyze the T-trajectory.


\par
The lower panels of Fig.~\ref{fig:figure1} compare the T-trajectory computed from Hamiltonian~\eqref{eq:main_Hamiltonian} and our model. The energy of the fictitious particle guiding the T-trajectory is denoted $\mathcal{E}_T$. If $\mathcal{E}_T >0$ (right panel), the guiding fictitious particle trajectory is unbounded, and the T-trajectory reaches the detector without recolliding, with final momentum $\mathbf{P}$. However, the T-trajectory is deflected due to the effective Coulomb interaction in the averaged Hamiltonian~\eqref{eq:averaged_Hamiltonian_H2}. The Coulomb asymmetry observed in Fig.~\ref{fig:figure2} is the direct consequence of this deviation. If $\mathcal{E}_T <0$ (left panel), the fictitious particle populates Rydberg states of Hamiltonian~\eqref{eq:averaged_Hamiltonian_H2} after tunnel-ionization, i.e., the guiding fictitious particle trajectory is bounded, and the electron must return to the ionic core. During rescattering, the energy of the guiding fictitious particle jumps to another energy level [since the averaged model~\eqref{eq:averaged_Hamiltonian_H2} is not valid close to the ionic core], and then could ionize if its energy is positive after rescattering. In addition, we notice that if the field envelope lasts only a few laser cycles, i.e., less than the period of the Rydberg orbit, then the electron is captured in a Rydberg state~\citep{Nubbemeyer2008}. 
\par
The energy of the fictitious particle $\mathcal{E}_T$ depends on the laser parameters, and in particular on the field ellipticity, through the change of initial coordinates~\eqref{eq:change_coordinates}. There exists a critical polarization $\xi_c$ such that $\mathcal{E}_T (\xi_c) = 0$. An approximation of the critical ellipticity is
\begin{equation}
\label{eq:critical_ellipticity}
\xi_c \simeq \dfrac{\sqrt{2}\omega^2}{E_0^{3/2}} \left( 1 + \gamma^2/2 \right)^{-1/2} ,
\end{equation}
where $\gamma = \sqrt{I_p/2\mathrm{U}_p}$ is the Keldysh parameter~\citep{Keldysh1965} and $\mathrm{U}_p = E_0^2/4\omega^2$ is the ponderomotive energy. We have assumed that $V(\mathbf{r}_g) \simeq -1/|\mathbf{r}_g|$, $\xi_c^2 \ll 1$, and $\mathbf{r}_0 \simeq I_p \sqrt{\xi_c^2 + 1}/E_0$. If $\xi < \xi_c$ then $\mathcal{E}_T < 0$, and the T-trajectory is rescattered. In our model, the observable $\mathbf{P}$ does not exist because the T-trajectory does not reach the detector. If $\xi > \xi_c$ then $\mathcal{E}_T > 0$, and the T-trajectory reaches the detector without recolliding. For $I = 8\times 10^{13} \unites{W\cdot cm}^{-2}$, the critical field polarization obtained from Eq.~\eqref{eq:critical_ellipticity} is $\xi_c \approx 0.32$, in agreement with upper panel of Fig.~\ref{fig:figure1}. For $I = 8\times 10^{14} \unites{W\cdot cm}^{-2}$, the critical field polarization obtained from Eq.~\eqref{eq:critical_ellipticity} is $\xi_c \approx 0.08$, in agreement with experimental measurements~\citep{Landsman2013}. For $I = 1.2 \times 10^{14} \unites{W\cdot cm}^{-2}$, a wavelength of $790 \unites{nm}$ and an $\mathrm{Ar}$ atom, it is given by $\xi_c \approx 0.27$, also in agreement with experimental measurements~\citep{Li2017}. We notice that the ad hoc criterion used in Ref.~\citep{Landsman2013} based on the perturbed SFA theory~\citep{Goreslavski2004} does not provide a correct estimate of $\xi_c$ for intensities smaller than $5 \times 10^{14} \unites{W\cdot cm}^{-2}$.
\par
The final momentum of the fictitious particle guiding the T-trajectory is $\mathbf{P} = \sqrt{2 \mathcal{E}_T}( \hat{\mathbf{x}} \cos\Theta + \hat{\mathbf{y}}  \sin\Theta)$ for $\xi \geq \xi_c$, where $\Theta$ is the scattering angle of the fictitious particle guiding the T-trajectory. Assuming that $V(\mathbf{r}_g) \simeq - 1/|\mathbf{r}_g|$, the scattering angle is $\Theta = \pi/2 +  \sin^{-1} ( 2 \mathcal{E}_T \ell^2 + 1 )^{-1/2}$, where $\ell$ is the guiding fictitious particle angular momentum. Close to the bifurcation, the guiding fictitious particle energy is $\mathcal{E}_T \approx 4 \mathrm{U}_p \xi_c (\xi - \xi_c)$, and we have
\begin{eqnarray}
P_x &\approx& - \sqrt{2 \xi_c} (E_0/\omega) (\xi - \xi_c)^{1/2} , \nonumber \\
P_y &\approx&  4 (E_0/\omega) (\xi - \xi_c) . \nonumber 
\end{eqnarray}
We notice that the bifurcation is observed for both $P_x$ and $P_y$. Consequently, we show that Coulomb focusing breaks down when Coulomb asymmetry becomes significant, as experimentally observed~\citep{Landsman2013}. 


\begin{figure}
\centering
\includegraphics[width=0.5\textwidth]{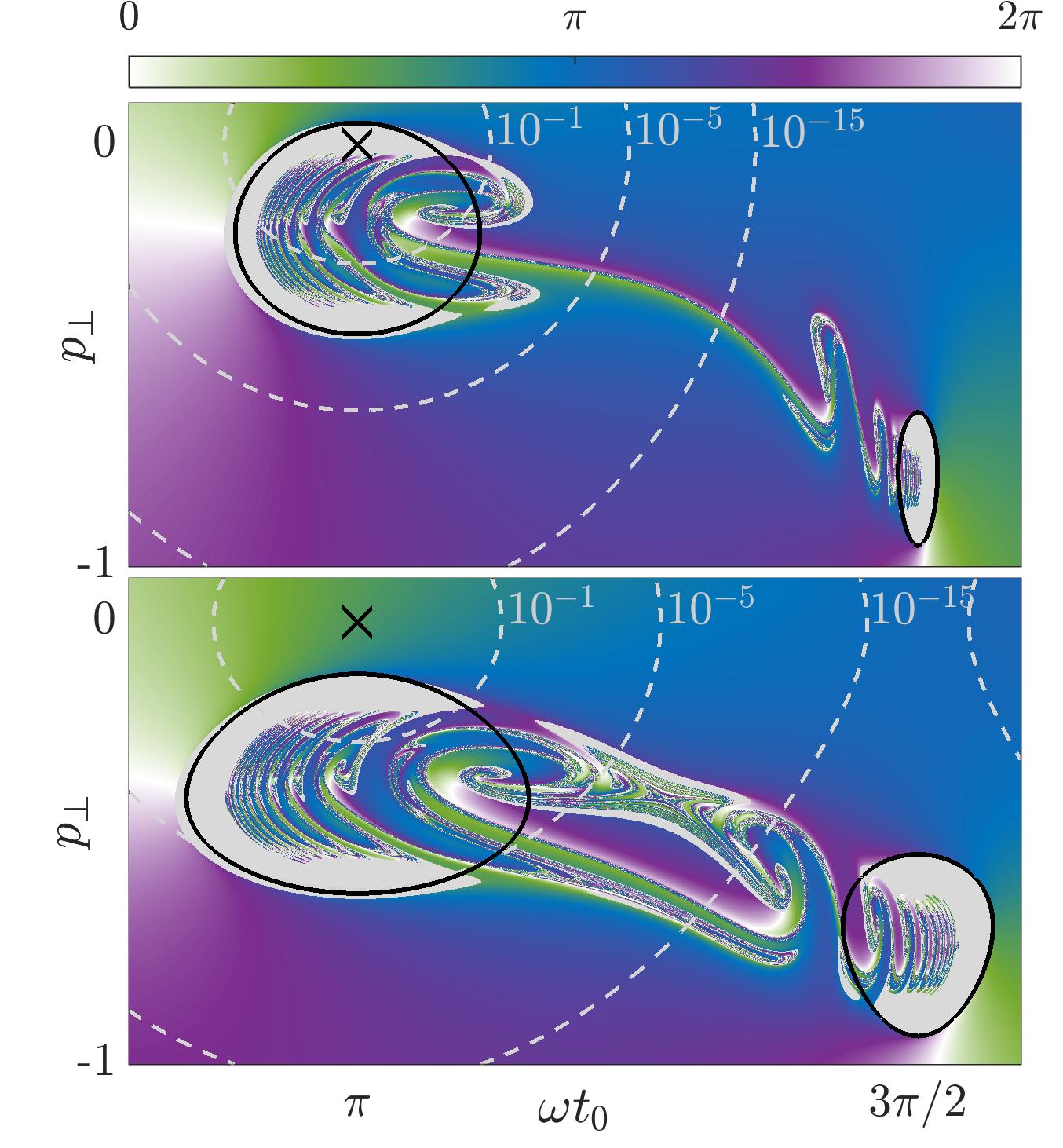}
\caption{Scattering angle of the electron of Hamiltonian~\eqref{eq:main_Hamiltonian} as a function of the initial conditions $(t_0, p_{\perp})$, for $\xi = 0.25$ (upper panel) and $\xi = 0.5$ (lower panel), for a field envelope $f$ with an eight laser-cycle plateau and a two laser-cycle ramp-down. The final electron energy is negative in grey areas. The black lines show where the guiding fictitious particle energy is $\mathcal{E} = 0$ according to our model~(\ref{eq:averaged_Hamiltonian_H2}). The crosses show the initial conditions of the T-trajectory. The momentum $p_\perp$ is in atomic units.}
\label{fig:figure3}
\end{figure}

\par
Two kinds of photoelectrons coexist --direct and rescattered electrons-- and contribute to the PMDs, and both are essential for probing the ion-electron interaction. However, the chaotic behavior of the rescattered electron trajectories, as shown in high-energetic part of ATI spectra~\citep{Walker1996}, reduces their local contribution in the PMDs. Figure~\ref{fig:figure3} shows the scattering angle of the electron as a function of the initial conditions $(t_0, p_{\perp})$, computed from the trajectories of Hamiltonian \eqref{eq:main_Hamiltonian}. We observe chaotic regions which are the signature of the highly nonlinear interactions driving the electrons during rescattering. Two main chaotic regions, centered at $\omega t_0 = \pi$ and $\omega t_0 = 3\pi/2$, are surrounded by initial conditions leading to electrons trapped into Rydberg states. We refer to this set of domains as the rescattering domain. In our model, the rescattering domain is determined by $\mathcal{E}<0$, and the black lines in Fig.~\ref{fig:figure3} are its boundaries $\mathcal{E} = 0$. We notice the very good agreement between the region ${\cal E}<0$ in our model and the set of trajectories which have undergone rescattering or remained trapped in Rydberg states. In the upper panel of Fig.~\ref{fig:figure3}, we observe that for $\xi < \xi_c$, the initial conditions of the T-trajectory belong to the rescattering domain. Hence, even if the rescattered trajectories are heavily weighted by the ADK ionization rate, their local contribution in the PMDs is relatively weak. Consequently, the electrons that contribute the most are the ones close to the boundaries of the rescattering domain, corresponding to electrons reaching the detector with energy $\mathcal{E} = 0$. Therefore, the maximum of the ATI spectrum is at zero energy. As the laser parameters are varied, particularly the ellipticity, the rescattering domain moves in the plane of initial conditions after tunnel-ionization. For $\xi > \xi_c$, the T-trajectory no longer belongs to the rescattering domain, as seen in the bottom panel of Fig.~\ref{fig:figure3}, so the ATI spectrum is peaked at $\mathcal{E}_T$ and the PMDs are dominated by direct electrons. Thus, we predict that the peak of the ATI spectrum is located at $\mathcal{E} = \max \lbrace 0, \mathcal{E}_T \rbrace$, as shown in the upper panel of Fig.~\ref{fig:figure1}. We notice that when $\xi$ increases further away from $\xi_c$, the rescattering domain moves to regions of initial conditions with very low ADK ionization rate. Consequently, the contribution of rescattered electrons and electrons with energy $\mathcal{E} = 0$ in the PMDs becomes very weak. Hence, we observe a lack of electrons in the neighborhood of the origin of the PMDs.

\par
In summary, we determined the microscopic mechanisms responsible for the shape of PMDs from the analysis of Hamiltonian~\eqref{eq:main_Hamiltonian}, and in particular, we showed that the change of shape observed in Ref.~\citep{Landsman2013, Li2017} as ellipticity is varied corresponds to a bifurcation. Our approach is based on a fictitious particle guiding the photoelectron motion. This model provides several predictions on the photoelectron motion and the shape of the PMDs, and allows the control of the ratio between the yield of rescattered and direct electrons, elements which are essential for imaging techniques.

We acknowledge Cornelia Hofmann and Ursula Keller for helpful discussions. The project leading to this research has received funding from the European Union's Horizon 2020 research and innovation program under the Marie
Sk{\l}odowska-Curie grant agreement No. 734557. S.A.B. and T.U. acknowledge funding from the NSF (Grant No.
PHY1602823).


%

\end{document}